# SMART-WRITE: Adaptive Learning-based Write Energy Optimization for Phase Change Memory

1st Mahek Desai
Computer Science Department
California State University, Northridge
Northridge, USA
mahek-trushit.desai.849@my.csun.edu

2nd Rowena Quinn
Computer Science Department
Illinois Central College
Peoria, USA
rowenaquinn1@gmail.com

3rd Marjan Asadinia
Computer Science Department
California State University, Northridge
Northridge, USA
marjan.asadinia@csun.edu

*Abstract*—As dynamic random access memory (DRAM) and other current transistor-based memories approach their scalability limits, the search for alternative storage methods becomes increasingly urgent. Phase-change memory (PCM) emerges as a promising candidate due to its scalability, fast access time, and zero leakage power compared to many existing memory technologies. However, PCM has significant drawbacks that currently hinder its viability as a replacement. PCM cells suffer from a limited lifespan because write operations degrade the physical material, and these operations consume a considerable amount of energy. For PCM to be a practical option for data storage—which involves frequent write operations—its cell endurance must be enhanced, and write energy must be reduced. In this paper, we propose SMART-WRITE, a method that integrates neural networks (NN) and reinforcement learning (RL) to dynamically optimize write energy and improve performance. The NN model monitors real-time operating conditions and device characteristics to determine optimal write parameters, while the RL model dynamically adjusts these parameters to further optimize PCM's energy consumption. By continuously adjusting PCM write parameters based on real-time system conditions, SMART-WRITE reduces write energy consumption by up to 63% and improves performance by up to 51% compared to the baseline and previous models.

## I. INTRODUCTION

As current transistor-based memory technologies like DRAM approach their scalability limits, there is a growing need for alternative storage solutions. Phase change memory (PCM) has emerged as a promising candidate due to its ability to use chalcogenide materials like $Ge_2Sb_2Te_5$ (GST) [1] to switch between low-resistive crystalline and high-resistive amorphous states for data storage. PCM offers advantages such as high scalability, non-volatility, low leakage power, and competitive read latency, positioning it as a leading contender for next-generation main memory systems [2]–[5]. PCM functions through two primary processes: SET and RESET operations [6]. During SET operations, the material is heated below its melting point and then slowly cooled to achieve a crystalline state, representing "1" in memory. Conversely, RESET operations involve heating the material above its melting point and rapidly cooling it to create an amorphous state, representing "0". These operations require significant electrical power due to the heating and cooling processes involved [6].

However, PCM faces significant challenges related to write energy, primarily because each write operation demands substantial energy to heat the memory cell and induce phase changes. This high energy requirement not only leads to increased power consumption but also generates considerable heat, which can affect system stability and necessitates effective thermal management. The intense energy involved in writes accelerates wear and degradation of the PCM cells, impacting their longevity (endurance). As a result, PCM cells can typically withstand only around $10^7$ to $10^9$ write operations before degradation occurs, leading to potential data retention issues [3], [5]. For applications requiring frequent write operations, this limitation can reduce reliability and durability.

Addressing these issues requires innovative solutions such as designing more energy-efficient write strategies, and developing advanced thermal management techniques to reduce energy consumption and enhance the overall durability and performance of PCM systems. To overcome these challenges, our proposed approach, SMART-WRITE, involves integrating neural network (NN) and reinforcement learning (RL) models. The NN model would predict key metrics like write energy, latency, and endurance based on parameters such as voltage, current, and pulse duration under varying conditions, including voltage fluctuations, change in ambient temperatures, etc. Meanwhile, the RL model would dynamically optimize PCM write parameters in real-time based on system conditions and device characteristics. This adaptive learning process aims to minimize energy usage and latency while extending PCM cell endurance, thereby enhancing its suitability as a next-generation memory technology. This innovative application of machine learning techniques represents a novel approach to overcoming critical limitations in PCM technology.

The remainder of this paper is organized as follows: Section II details the proposed method, including the integration of a neural network (NN) model for predicting write parameters in PCM, as well as the structure of the reinforcement learning (RL) model we will implement to optimize write energy and latency. Section III discusses the evaluation of our results, illustrating the simulator we will be implementing to gather our data and compare our results, and highlighting the accuracy of our NN and the decrease in total energy, write energy and latency we were able to achieve utilizing the RL model. Section IV explores the comparison of SMART-WRITE with existing methods such as MCT [7] and RRM [13]. Finally, section V concludes the paper, summarizing our contributions and the potential impact of our proposed methods on the development of PCM as a next-generation memory technology.

## II. PROPOSED METHOD

In this section, we explain our proposed SMART-WRITE approach in detail. We begin with a discussion on collecting data about the write process, followed by an explanation of data preprocessing techniques to clean and prepare the data for machine learning model training.

We then describe the development of adaptive learning algorithm for optimizing write processes by training machine learning models with the collected data. The flow between each section is visualized in Figure 1.

### A. Data Collection

*1) Parameter Generation:* In order to optimize write energy and latency, we generated parameters for the NVMain simulator [11], [12] focusing on various voltage levels and pulse durations for both set and reset operations, as well as varying read:write ratios. NVMain simulator is widely recognized in academic research as a reliable and detailed simulator for emerging non-volatile memory technologies [7], [14]–[17]. In our work with NVMain, we use parameters derived from real PCM devices and prior research to ensure that our simulation results closely reflect realistic device behavior. We have also integrated real-world constraints, including temperature effects, process variations, and realistic endurance models, into our simulations. These efforts help bridge the gap between the simulated environment and actual device performance.

For the set parameters, the voltage is treated as a discrete variable of values 1.5 V, 2.0 V, and 2.5 V, while the pulse duration is considered with discrete values of 150 ns, 155 ns, and 160 ns. For the reset parameters, the voltage is a discrete variable of values 2.5 V, 3.0 V, and 3.5 V, and the pulse duration includes discrete values of 100 ns, 105 ns, and 110 ns, along with ambient temperature with a discrete variables of values 25°C, 50°C, and 75°C. We used the following range for these variables since we obtained similar results with different ranges as well, ensuring robustness and consistency in our findings.

Additionally, we generated trace files serving as input to the NVMain simulator, using varied read:write ratios, including 9 : 1, 8 : 2, 7 : 3, and so on. These trace files were generated based on real application data (like SPEC CPU and PARSEC [20], [21]) to closely represent actual workload patterns. By capturing typical data access behaviors and diverse operational characteristics, we ensured the trace files realistically simulate real-world scenarios, enhancing the reliability of our evaluations in NVMain. Specifically, we prepared 20 trace files for scenarios where the read operations are greater than the write operations ($R > W$), 20 files for balanced operations ($R = W$), and 20 files for scenarios where the read operations are fewer than the write operations ($R < W$). This comprehensive dataset was created by combining and permuting all these values, covers a wide range of operational conditions.

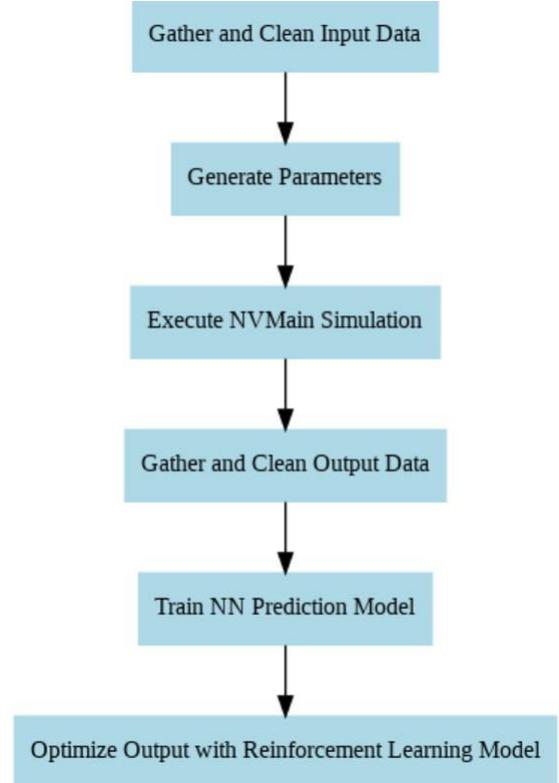

Fig. 1. SMART-WRITE Logic Block.

*2) Simulation Process:* The simulation process for running NVMain begins by reading the first row from the generated dataset file, which contains the parameter combinations. Next, we update the parameters such as set voltage, reset voltage, set pulse duration, and reset pulse duration in NVMain based on the values read. NVMain is then executed using the specified trace file corresponding to the parameter set. The simulation runs until 100,000 operations are completed, which is around 7.5 million simulation cycles. The results of the simulation are outputted into a designated folder. After the simulation and output generation, we move to the next row in the dataset and repeat the process until all rows are processed.

*3) Data Preprocessing:* After running the simulation, we collected the raw output data corresponding to the inputted values in NVMain. We then cleaned this raw data to extract the required values such as write energy, total energy, write latency, total latency, and endurance.

In preparing the data for the predictive model, we undertook several preprocessing steps. We One-Hot-Encoded all discrete input variables in order to avoid ordinality when training our model. After this process we had 14,580 rows of data, which we split into training, testing, and validation sets. 60% of the data was devoted to training the model, 20% was devoted to testing, and 20% to validation. This ensured that when evaluating the model, we could determine if the neural network was robust enough to adapt to new data and not over-trained on the simulation data.

*B. Static Model*

In order to create an RL model that dynamically adjusts to the operating conditions of the system, we first needed to create a static model that could predict the metrics we want to optimize (like write energy, write latency, and endurance). After gathering data from NVMain, we would separate the data into our inputs (features) and outputs (targets). We decided on voltage and pulse duration of set and reset operations as well as the total reads and writes that occurred in a trace file and ambient temperature as our features. The targets would be what we want to optimize: total write energy, total write latency, and endurance. We then built a model using a Multi-Layer Perceptron (MLP) architecture with these input features and output targets.

The MLP Neural Network (NN) was built as a multi-output regression model utilizing the Functional API structure of the Keras library [8] and tuned using the Keras Tuner [8]. Through examination of NVMain's method of calculating write energy, write latency, and endurance, we were able to separate the internal structure of the model so that only relevant inputs were used to train the model on the respective outputs.

The output features with their respective input features are shown in Table I.

TABLE I
INPUT FEATURES WITH RESPECTIVE OUTPUT FEATURES.

| Input Features | Output Features |
|---|---|
| SET/RESET Voltage<br>SET/RESET Pulse Duration<br>Read:Write Ratio<br>Ambient Temperature | Total Write Energy |
| SET/RESET Pulse Duration<br>Read:Write Ratio<br>Ambient Temperature | Total Write Latency |
| Read: Write Ratio<br>Ambient Temperature | Endurance |

This resulted in a model that took multiple features and separated the features to feed into and pass through 7, 5, and 5 dense layers to predict the respective targets (write energy, write latency, and endurance).

Tuning was implemented using the Keras tuner [8] for the following hyperparameters: number of hidden layers, number of neurons per hidden layer, the loss function, the L1 and L2 values of the kernel regularizer, batch size, and the regression optimizer. Hyperparameters are tunable values during training so, the tuner then ran through multiple combinations of different values of each aspect until it found the combination that gave the least mean absolute error (MAE), and validation loss. The NN overall landed on the optimizer Adam for write latency and endurance models and Nadam for write energy model, the loss function Huber for all three, and a batch size of 160 for write latency and endurance models and 384 for write energy model. The resulting differences in model structure when the NN separates into different dense layers can be shown in Table II, and a simple diagram of the MLP NN can be shown in Figure 2.

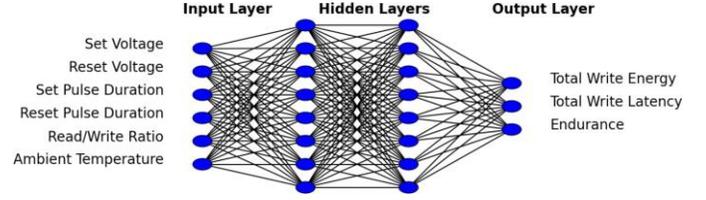

Fig. 2. MLP Neural Network Structure with Input layer, Hidden layers and Output layer.

When training the model, two callbacks were utilized to optimize the performance of the training, ensuring minimal validation loss. The learning rate of the adam optimizer was initialized at 0.001 and decreased by 0.1 whenever the validation loss would plateau for three epochs, and the fitting would stop once the validation loss had not improved for four epochs. This ensured that the model would not over-train.

TABLE II
NEURAL NETWORK STRUCTURE PER OUTPUT.

| Model Output | Hidden Layers | Neurons per Layer | Kernel Regularization L1, L2 Weights |
|---|---|---|---|
| Write Energy | 7 | 8,20,8,12,32,32,40 | 0.001,0.1 |
| Write Latency | 5 | 30,14,24,16,12 | 0.01,0.001 |
| Endurance | 5 | 30,14,24,16,8 | 0.01,0.001 |

*C. Dynamic Model*

Once we had a NN that could predict the output features, we could begin constructing a Reinforcement Learning (RL) process that could dynamically adjust the SET and RESET pulse duration and voltage in order to optimize the total write energy and latency without adversely affecting endurance that occurs during PCM operation. The details of our RL process are listed as follows:

**Custom Environment:** Using the Gymnasium library [9], we developed a custom RL environment specific to PCM operations, incorporating write energy, latency, and endurance calculations. The action space, represented by a Multi-Discrete space, included choices for SET/RESET voltage and pulse duration values. This resulted in an action space with four actions, each with three discrete options. In each step, the agent performed a write or read operation using a randomly generated binary trace. Write energy, latency, and endurance metrics were recorded for each write operation. A trained NN model then evaluated these metrics, providing feedback to the RL agent.

**Action Space:** The action space is represented in the following manner: Within the step function, the agent would perform an action by selecting a SET voltage and pulse duration and a RESET voltage and pulse duration. These actions were contained within a Multi-Discrete action space, as each action was a choice between different discrete values. This resulted in an action space that contained four actions with three possible values each. A random binary trace was then generated, and a write or read operation would occur. The energy, latency and endurance of the operation was recorded if it was a write, and then the RL would reward the agent utilizing the NN we previously trained.

**Reward Function:** Within the reward function, the NN would take the current action state and predict the total write energy and total write latency as if the model would continue to use the current values for PCM operation. If the predicted values were less than the previous step's predicted values of write energy and write latency and more than that of endurance then the model would give the agent a minor reward (+0.25), and if the predicted values were more for write energy and write latency and less for endurance then the agent receives a minor punishment (-0.25). This would ensure that the agent was making steps in the right direction (less write latency and write energy without significantly affecting endurance). If the predicted values of write energy and write latency were less than and that of endurance was more than a normal operation of PCM, then the agent receives a major reward (+10). This was our overall goal for the project, which meant that if the agent was able to achieve it, we wanted to ensure that it stuck with this action. After rewarding the agent, the model would then move on to the next step. This cycle would continue until 100,000 operations, which take place over 100,000 steps, were completed. This mimics a normal run of NVMain.

**RL Algorithm:** A simple diagram summarizing our reinforcement learning process is presented in Figure 3. The agent was trained using the Proximal Policy Optimization (PPO) algorithm with an MLP policy that acted as an actor-critic from Stable-Baselines3 [10]. Unlike Q-learning approaches that learns to approximate how valuable a given state is, PPO optimizes a policy that matches states to actions, effectively guiding adjustments in SET and RESET parameters. The PPO hyperparameters were set to default values in the Stable-Baselines3 library, shown in Table III.

**Training Process:** The RL agent interacts with the environment for up to 100,000 operations, where it explores different combinations of SET/RESET parameters. The agent updates its policy after each step based on the feedback (rewards), learning the optimal settings to minimize write energy and latency while maintaining endurance. The hyperparameters of the PPO were kept as the base values assigned within stablebaselines3 [10], shown in Table III.

**State:** The state represents the current operational conditions of the Phase Change Memory (PCM) system. In our work, this would include the parameters related to SET and RESET operations, such as voltage, pulse duration, ambient temperature and current read:write ratio. These parameters determine the current behavior of the memory cell during a write operation.

**Action:** The action is the decision that the RL agent makes in response to the current state. In the our model, the action space involves adjusting the SET voltage and pulse duration and RESET voltage and pulse duration to optimize the performance of the PCM. These adjustments aim to minimize write energy and latency without substantially minimizing endurance, ensuring that the RL model finds the best parameters for these operations.

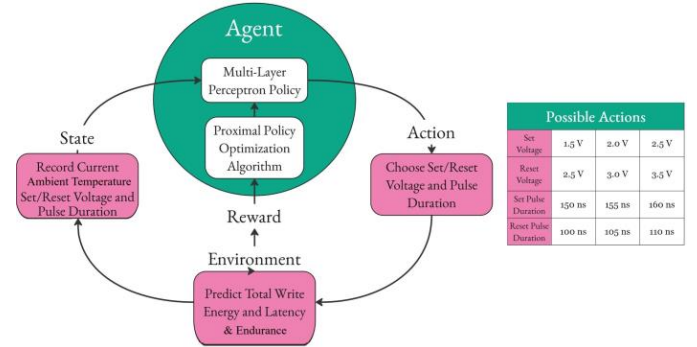

Fig. 3. PPO Reinforcement Learning Model.

TABLE III
PROXIMAL POLICY OPTIMIZATION ALGORITHM (PPO)
HYPERPARAMETERS

| Parameter | Value |
| --- | --- |
| Learning Rate | 0.0003 |
| Batch Size | 64 |
| Discount Factor | 0.99 |
| Trade-Off Factor | 0.95 |
| Entropy Coefficient | 0.0 |
| Value Coefficient | 0.5 |

## III. EVALUATION

In this section, we describe our simulation environment and methodology, and present the evaluation results of the proposed method compared to the PCM baseline.

### A. Simulation Environment

We implemented our proposed method in NVMain [11], [12] which is a versatile, cycle-accurate memory simulator that models both conventional DRAM and new non-volatile memory (NVM) technologies like phase change memory (PCM). It offers detailed simulations of memory timing, energy consumption, and specific NVM traits such as limited write endurance and multi-level cells. Additionally, NVMain supports hybrid memory systems, fine-grained bank/subarray-level parallelism, and allows for custom memory controllers and address mapping schemes. For PCM, NVMain effectively models key aspects like asymmetric read/write latencies, write energy, and cell endurance, making it suitable for evaluating PCM optimization strategies [11], [12].

TABLE IV
CONFIGURATION ENVIRONMENT

| Parameter | Value |
|---|---|
| **Memory Specifications** | |
| Technology node | 20nm |
| Operating voltage | 1.8V |
| Device capacity | 8Gb |
| Program bandwidth | 40MB/s |
| **Interface Specifications** | |
| Clock frequency in MHz | 400 |
| Bus width in bits | 64 |
| Number of bits per device | 8 |
| CPU frequency in MHz | 2000 |
| **MLC Parameters** | |
| Number of MLC levels | 2 |
| **Memory Controller Parameters** | |
| Memory controller type | FRFCFS |
| Address mapping scheme | R:RK:BK:CH |
| Read queue size | 32 |
| Write queue size | 32 |
| **Endurance Model Parameters** | |
| Endurance model type | BitModel |
| Endurance distribution type | Normal |
| Endurance distribution mean | 1000000 |
| Endurance distribution variance | 100000 |

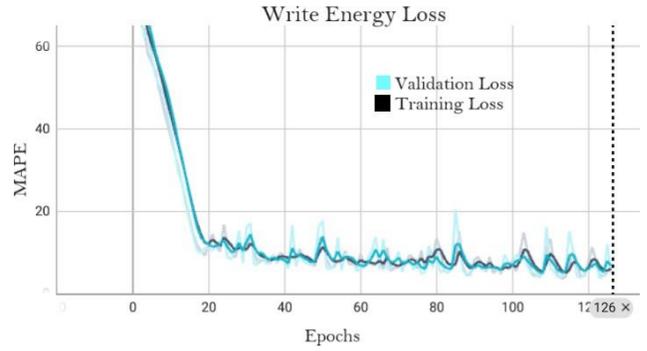

Fig. 4. Write Energy Loss Result.

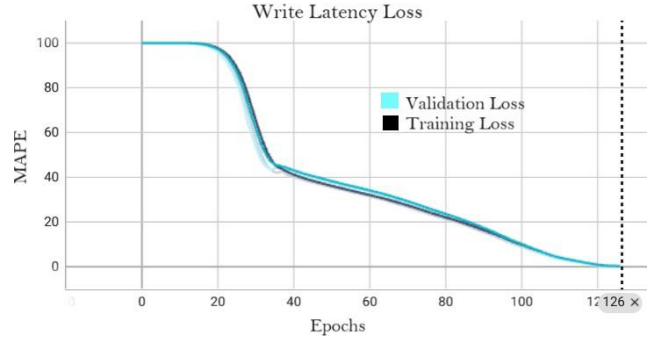

Fig. 5. Write Latency Loss Result.

*1) NVMain Architecture:* The simulation setup is outlined in Table IV. The setup includes a 20nm 1.8V 8Gb PRAM with 40MB/s program bandwidth. It also includes parameters such as clock frequency , bus width, CPU frequency, etc. These parameters were kept unchanged in the default configuration file in NVMain to mimic PCM as accurately as possible. These parameters are essential for accurately simulating and assessing the performance and energy consumption of the memory system under various conditions.

*B. Static Model Evaluation*

Since the NN's task was regression, accuracy is not as simple as looking at the proportion of correct predictions of the model. Consequently, to evaluate the performance of the NN we focused on the Mean Absolute Percentage Error (MAPE). The MAPE is the mean of the absolute differences between the actual values and the predicted values divided by the actual values. This gives us a percentage of how close the actual values are to the predicted values, with lower percentages relaying that the values are closer together.

When training the NN, as discussed in Section II-A3, we separated the data into different sets in order to perform training, testing, and validation. For training we set aside 60% of the data, and for testing and validation we set aside 20% each. We first evaluated how well the model fits the training data, then how well the model fits new data to ensure that the model would be adaptable to different datasets. This required us to evaluate the MAPE of the training loss (how well the model fits to training data) and the MAPE of the validation loss (how well the model fits to new data).

The results related to the trends of loss for both validation and training data, concerning energy, latency, and endurance, are shown in Figures 4, 5 and 6

In Figures 4, 5, and 6, you can see that the trends of loss for both validation and training data decreased drastically, and then were incrementally decreased by the learning rate suppression callback implemented into the model fitting. Since the validation loss follows the training loss closely, we can infer that the model did not overfit on the training data.

Once we were sure the model was not overfitted, we were able to move on to evaluating the accuracy of the NN. We evaluated the testing data utilizing MAPE as a metric, and were able to achieve the results shown in Table V.

We then plotted the regression graphs shown in Figures 7, 8, and 9, illustrating how far the predicted values stray from the actual values.

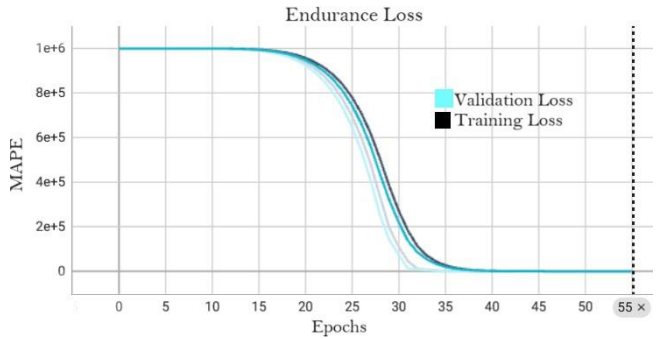

Fig. 6. Endurance Loss Result.

TABLE V
MAPE PER OUTPUT

| Output | MAPE (%) |
|---|---|
| Endurance | 0.00098 |
| Total Write Latency | 0.31 |
| Total Write Energy | 0.91 |

TABLE VI
RL REWARD STATS.

| Ratio | Mean Reward | SD Reward |
|---|---|---|
| Reads > Writes | 102.7 | 24.227 |
| Reads = Writes | 8730.8 | 36.43 |
| Reads < Writes | 18463.1 | 38.13 |

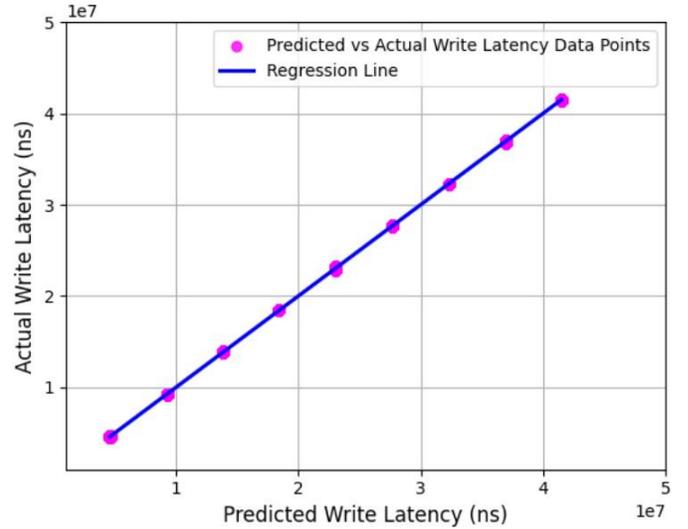

Fig. 8. Predicted vs. Actual Write Latency Regression.

### C. Dynamic Model Evaluation

Once we ensure the neural network (NN) has high accuracy, we can move on to implementing the reinforcement learning (RL) model. We tailored the model to gauge its performance under three different scenarios: 1) when the number of reads is greater than the number of writes, 2) when the number of writes is greater than the number of reads, and 3) when the number of reads and writes are equal. We then trained and evaluated the reinforcement learning model under each of these scenarios with the intention to compare its calculated total write energy to the PCM baseline model and NVMain's calculations.

In Table VI, the mean and standard deviation (SD) of the rewards given to the agent each evaluation cycle is listed. Since the mean reward is positive, we can tell that the agent was successful in finding a good strategy to optimize write energy and latency without substantially impacting endurance.

The following Figures 10, 11, and 12 illustrate the performance comparison between our proposed SMART-WRITE method and the baseline model in terms of key metrics such as write energy, total energy, and write latency respectively for ambient temperatures 25°C, 50°C and 75°C. Unlike traditional approaches that maintain constant voltage values irrespective of ambient temperature, our adaptive model allows the voltage to decrease with rising ambient temperatures.

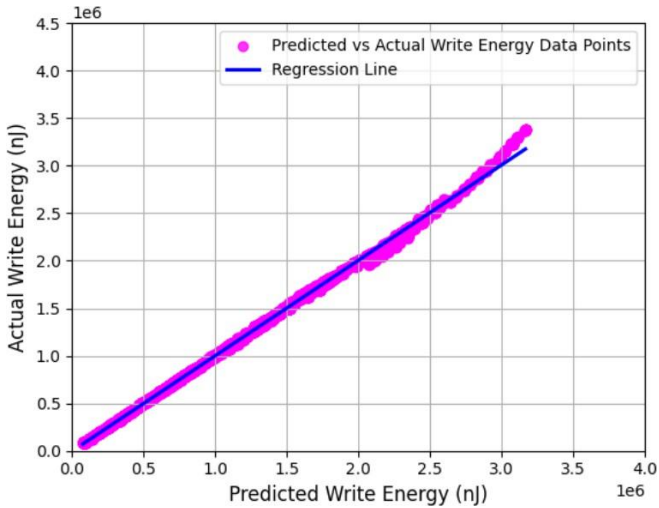

Fig. 7. Predicted vs. Actual Write Energy Regression.

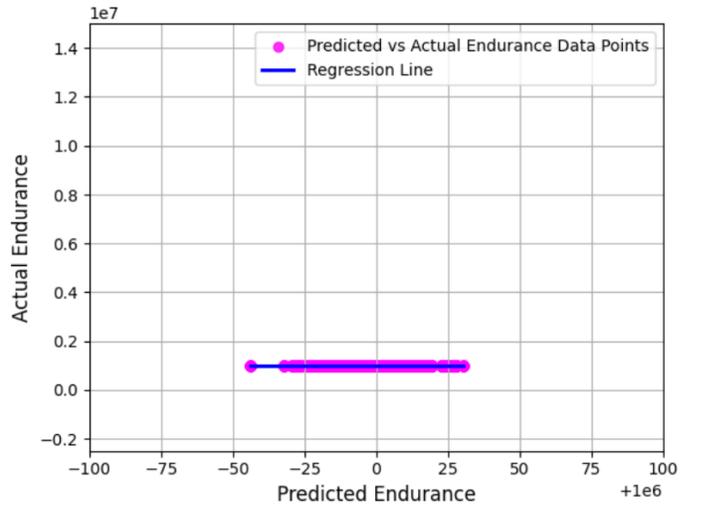

Fig. 9. Predicted vs. Actual Endurance Regression.

It is notable that, to model the temperature-dependent behavior of voltage in PCM, we define the voltage, $V_{set/reset}(T)$, as a function of the ambient temperature $T$ as seen in Equation 1. The derivation of Equation 1 is based on recent research findings on the thermal characteristics and switching mechanisms of phase-change memory (PCM). As observed in [18], PCM's thermal resistance varies significantly with ambient temperature, affecting power density. This effect is particularly notable in confined PCM cells where the relationship between temperature and power is inversely proportional. By acknowledging this thermal resistance dependency, we can ap-proximate that higher ambient temperatures reduce the power (and, consequently, the voltage) required for set and reset operations due to the decreased thermal resistance. Adding to it, in [19], phase transitions in PCM are influenced by temperature changes due to the required activation energy for state change. The temperature coefficients $\alpha_{set}$ and $\alpha_{reset}$ represent these sensitivities in the empirical model, describing how temperature directly affects the set/reset voltage thresholds necessary for stable phase transitions. Their findings reinforce that a linear approximation can capture this temperature effect on reset voltage, thus supporting our formula.

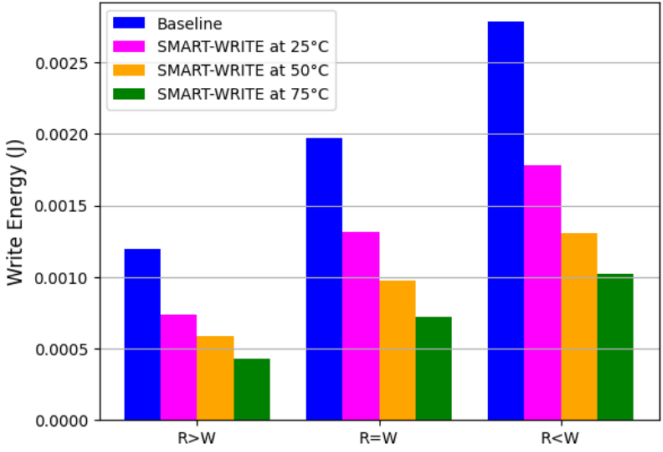

Fig. 10. Write Energy, SMART-WRITE vs Baseline.

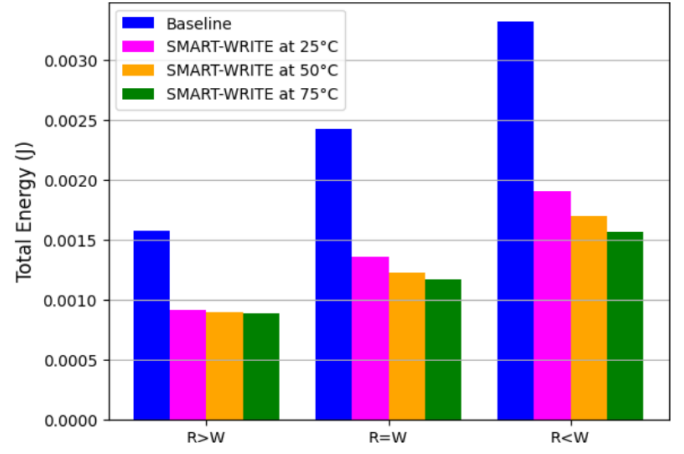

Fig. 11. Total Energy, SMART-WRITE vs Baseline.

$$V_{set/reset}(T) = V_{set/reset}(T_0) - \alpha_{set/reset} \cdot (T - T_0) \quad (1)$$

where:

- $V_{set/reset}(T_0)$ is the set/reset voltage at room temperature $T_0 (= 25°C)$.
- $\alpha_{reset}$ represents the temperature coefficient (= 0.015) for voltage in reset operation [19].
- $\alpha_{set} = 0.025$ is temperature coefficient for voltage in set operation [19].

The results in the following demonstrate how the integration of reinforcement learning (RL) into PCM systems can significantly optimize both energy efficiency and performance. The improvements achieved by SMART-WRITE highlight its potential as an adaptive solution for enhancing the overall operation of phase-change memory. Figure 10 highlights the comparison between the SMART-WRITE method and the baseline model in terms of total write energy consumption. The results clearly show that SMART-WRITE reduces write energy by as much as 63% across varying ambient temperatures. This significant reduction demonstrates the effectiveness of the reinforcement learning (RL) approach in optimizing the parameters of the phase-change memory (PCM) system, making it much more energy-efficient compared to the traditional baseline approach.

In Figure 11, we compare the total energy consumed by the system, incorporating both write and read operations, for SMART-WRITE and the baseline model. Similar to write energy, SMART-WRITE achieves a substantial reduction in total energy consumption, up to 52% across varying ambient temperatures. This showcases the adaptability of the proposed model in optimizing the energy use of PCM, not just for write operations but for overall memory performance.

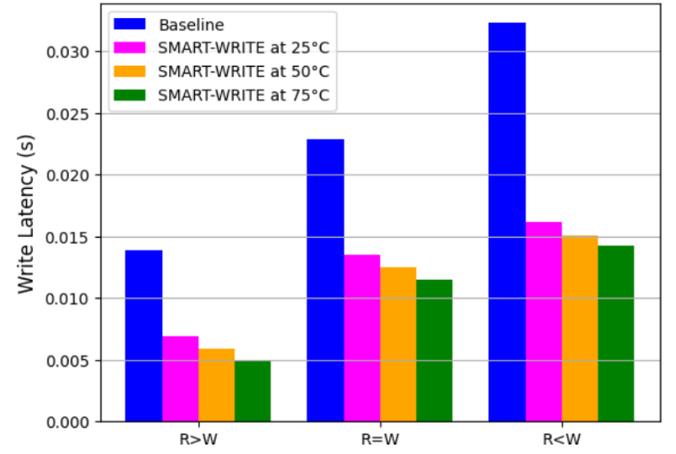

Fig. 12. Write Latency, SMART-WRITE vs Baseline.

Figure 12 focuses on the write latency comparison between SMART-WRITE and the baseline. SMART-WRITE shows about 51% improvement in write latency, demonstrating its ability to optimize not only energy consumption but also performance. By dynamically adjusting write parameters, the reinforcement learning model reduces the time it takes to perform write operations, leading to faster memory access and improved system efficiency.

In summary, the results show consistent gains in efficiency for write energy, total energy, and write latency. Write energy and total energy consumption decreases dynamically with temperature as seen in Figures 10 and 11 respectively, and latency is minimized due to the reduced voltage requirement as shown in Figure 12. These benefits make SMART-WRITE particularly advantageous for devices deployed in environ-ments with significant temperature fluctuations, enabling consistent performance and improved longevity. By continuously adapting PCM's operational parameters in response to ambient conditions, the proposed model not only achieves significant energy savings but also optimizes latency without affecting memory lifetime. These results establish SMART-WRITE as a viable solution for the energy-efficient, high-performance operation of PCM systems in modern applications.

## IV. SMART-WRITE COMPARISON WITH EXISTING METHODS

We compared the performance of our SMART-WRITE approach with two existing methods: Memory Cocktail Therapy (MCT) [7] and Region Retention Monitor (RRM) [13]. MCT is a machine learning-based framework that optimizes the performance, energy efficiency, and lifetime of Non-Volatile Memory (NVM) technologies like PCM by balancing trade-offs between performance, energy, and endurance. It operates by selecting the most impactful features, such as write latencies and cancellation options, and predicts performance and energy consumption using models like quadratic regression and gradient boosting. MCT showed improvements in both energy efficiency and performance. However, MCT's performance varies based on the memory technology and applications used. While MCT provides significant benefits, the complexity of managing a large configuration space remains a challenge, even with dimensionality reduction.

Similarly, the RRM [13] approach addresses the trade-off between write latency and retention in MLC PCM by dynamically adjusting write operations based on memory region access patterns. RRM identifies frequently written regions and applies faster, low-retention writes to improve performance, while selectively refreshing these regions to prevent data loss. Infrequently accessed regions receive slower, high-retention writes, which conserves memory endurance. RRM achieves a balance between performance and memory longevity, extending memory lifetime.

Table VII summarizes the performance of SMART-WRITE in comparison to the Memory Cocktail Therapy (MCT) and Region Retention Monitor (RRM) approaches. The table highlights three key metrics: total energy, performance, and endurance.

**Total Energy:** SMART-WRITE achieves a 52% reduction in total energy consumption, significantly surpassing MCT's 7.95% improvement. This is due to SMART-WRITE's adaptive learning approach, where reinforcement learning (RL) dynamically adjusts SET/RESET voltages and pulse durations in real time. By continuously optimizing these parameters based on system conditions, SMART-WRITE minimizes energy expenditure during write operations, offering superior efficiency compared to the static configurations of MCT and RRM.

While both SMART-WRITE and MCT use machine learning to optimize PCM, SMART-WRITE provides more dynamic and granular control. MCT relies on pre-sampled configurations and feature selection but struggles with managing a large configuration space. In contrast, SMART-WRITE's predictive model, paired with RL, continuously adapts to system conditions, making it more responsive and efficient, particularly in reducing write energy. RRM, meanwhile, focuses on managing write latency by classifying memory regions as "hot" or "cold" based on access patterns. However, RRM is primarily reactive, while SMART-WRITE proactively optimizes energy consumption and memory endurance through real-time predictions and dynamic adjustments.

**Performance:** In terms of performance, SMART-WRITE delivers a 51% improvement, which is much higher than MCT's 9.24% gain and comparable to RRM's 62% boost achieved through short writes. The primary reason for this performance gain is SMART-WRITE's ability to adaptively tune write parameters using reinforcement learning, ensuring that write latency is minimized without sacrificing accuracy or energy efficiency. This real-time optimization allows for a more balanced approach, improving both energy consumption and write speed in comparison to MCT's predefined configuration adjustments and RRM's static region-based strategy.

TABLE VII
SMART-WRITE COMPARISON WITH MCT AND RRM.

| Comparison Metric | SMART-WRITE | MCT | RRM |
|---|---|---|---|
| Total Energy | Up to 52% improvement. | 7.95% improvement. | 32.8% more energy consumption than Static-7-SETs (deterioration). |
| Performance | Up to 51% improvement. | 9.24% performance gain. | 62% performance improvement. |
| Endurance | Up to 1% improvement. | Uses techniques like Wear Quota to ensure lifetime targets. | Achieves a lifetime of 6.4 years (compared to 10.6 years for long writes and 0.3 years for short writes). |

**Endurance:** Although SMART-WRITE's 1% endurance improvement might seem modest compared to RRM's approach, which extends memory lifetime to 6.4 years, the main reason for this is that SMART-WRITE prioritizes optimizing energy and performance while still maintaining endurance. Unlike RRM, which relies on specific region-based refreshes to extend lifespan at the cost of higher energy consumption, SMART-WRITE focuses on reducing the overall energy impact on the PCM cells, indirectly preserving their longevity.

The overall improvements in energy and performance demonstrated by SMART-WRITE are primarily due to its dynamic, adaptive learning model that continuously optimizes parameters in real-time. This approach enables SMART-WRITE to effectively balance energy consumption, performance, and endurance in a way that static or semi-adaptive models like MCT and RRM cannot achieve.

## V. Conclusion

This paper introduces an innovative approach to optimize PCM write energy consumption and performance using neural networks (NN) and reinforcement learning (RL). By dynamically optimizing write parameters based on real-time operational conditions and device characteristics, our proposed method aims to reduce write energy consumption and latency while improving PCM endurance. By reducing write energy consumption and latency within PCM, memory operations within computers utilizing this technology will consume less power, which will greatly reduce the energy consumed by these devices over the course of their lifetime.

Initial findings from our evaluation involved comprehensive parameter generation and simulation processes using NVMain. Our predictive NN model was able to achieve high accuracy when predicting key output metrics, displaying the viability of predicting the behavior of PCM when given the system's parameters. The RL model was able to achieve a reduction of up to 63% of the total write energy and a performance improvement of up to 51% without adversely affecting the endurance across various process variation scenarios compared to baseline models. By utilizing the reinforcement learning (RL) model with the incorporation of the thermal model, we can improve the feasibility of phase-change memory (PCM) as a viable commercial main memory technology.

## VI. Acknowledgment

This work was supported by the National Science Foundation (NSF) under Grant Nos. [CNS-2244391] and [2318553]. We gratefully acknowledge the NSF for their financial support through both grants and for providing the resources necessary to conduct this research. The views expressed in this paper are those of the authors and do not necessarily reflect the views of the NSF.